\documentclass[aps,prl,twocolumn,floatfix]{revtex4}
\usepackage{epsfig,epsf}
\usepackage{amsmath}
\usepackage{graphicx}

\def\){\right)}
\def\({\left(}
\def\]{\right]}
\def\[{\left[}

\newcommand{\be}{\begin{equation}}
\newcommand{\ee}{\end{equation}}

\newcommand{\roughly}[1]%
{\mathrel{\raise.4ex\hbox{$#1$\kern-.75em\lower1ex\hbox{$\sim$}}}}

\newcommand\beq{\begin{eqnarray}}\newcommand\eeq{\end{eqnarray}}

\def\Dsl{\,\raise.15ex \hbox{/}\mkern-12.8mu D}

\def\fm3{fm$^{-3}$}

\begin{document}

\title{Superfluid-Normal Quantum Phase Transitions in an Imbalanced Fermi Gas}

\author{Heron Caldas}

\affiliation{Departamento de Ci\^{e}ncias Naturais, Universidade Federal de S\~{a}o Jo\~{a}o Del Rei, Pra\c{c}a Dom Helv\'{e}cio 74, 36301-160, S\~{a}o Jo\~{a}o Del Rei, MG, Brazil \\ }

\begin{abstract}

We investigate the superfluid-to-normal zero temperature quantum phase transitions of asymmetric two-component Fermi gases as a function of the chemical potential imbalance $h$. The calculations are performed for homogeneous and trapped imbalanced systems. We concentrate at unitarity, characterized by a divergent interaction parameter $k_F a$, where most of the current experiments are realized. For homogeneous systems, we determine the critical chemical potential imbalance $h_c$ at which possible phase transitions occur. In the case of trapped gases, we show how $h_c$ can be consistently determined from experimental observations.

\end{abstract}
\maketitle

\section{Introduction}

The extraordinary experimental advances in the physics of ultracold atoms in the last few years, has given to these systems the possibility to serve as a {\it laboratory} for the investigation (and simulation) of several theoretical ideas of physics, from condensed matter to high energy physics~\cite{Lewenstein}. Indeed, the recent observation of superfluidity in ultracold Fermi gases paved the way between atomic and solid state physics~\cite{Review1,Review2}.

Fermionic particles with different spin configurations, occupying states with equal and opposite momenta close to their common Fermi surface, form pairs. This is explained by the well known BCS theory of superconductivity. The presence of spin imbalance prevents this mechanism, since there are two Fermi surfaces now, that do not coincide and pairing with zero total momentum are disadvantageous to occur~\cite{Teo1}. At zero temperature $T$ and for small asymmetries between the two spin species, the system is still a superfluid. However, when the imbalance between the two Fermi surfaces is too large, superfluidity is broken apart and the system undergoes a quantum phase transition toward a normal state. The existence of such a transition at a critical value of the polarization was first proposed by Clogston~\cite{Clogston} and Chandrasekhar~\cite{Chandrasekhar}, who predicted the occurrence of a first-order phase transition from the superfluid to the normal state. This is acknowledged as the CC limit of superfluidity, and was originally conjectured in the context of conventional superconductivity, as {\it ``the maximum field at which a superconductor maintains superconductivity''}~\cite{Chandrasekhar}. The CC transition takes place when the polarization is large enough such that the normal phase becomes energetically more favorable than the superfluid state.

Experimental investigations with ultracold imbalanced Fermi gases, where the number of atoms in the two spin states is different, have shown that there can be more than (only) the distinct superfluid and normal phases. The first order transition between the superfluid at equal spin population and the imbalanced normal mixture brings about a phase separation between coexisting normal and superfluid phases in three-dimensional (3d)~\cite{Bedaque2003,Caldas2004} and two-dimensional (2d) systems~\cite{Souza,Thomas,Mitra}. Recent experiments using tomographic techniques, have found a sharp separation between a superfluid core and a partially polarized normal phase~\cite{PhaseSep}.

Thus, the exploration of a two-component Fermi gas with imbalanced populations is an active area of research in the field of ultracold atoms from both theoretical~\cite{Teo1,Teo2,Teo3,Teo4} and experimental points of view~\cite{Exp1,Exp2,Exp3,Exp4,Exp5}, which has given the opportunity for testing old and new ideas. As examples, we point out the search for the elusive FFLO phase, and for a connection between quantum phase transitions and quantum information~\cite{Lewenstein,QI}, respectively.

In this work we investigate the main zero temperature (quantum) superfluid-normal phase transitions that may occur in Fermi gases with imbalanced spin populations. We study the phase transitions that happens induced by an increase in the chemical potential imbalance $h = (\mu_{\uparrow} - \mu_{\downarrow})/2$, both in homogeneous and trapped 3d systems, and focus on the universal regime, at which the dimensionless parameter $k_F a$ diverges. We find the critical chemical potential imbalance responsible for the phase transitions from the superfluid to the normal-mixed and from the normal-mixed to a fully polarized phase and finally, from the fully polarized phase to the vacuum, for homogeneous and non-homogeneous systems. For the non-homogeneous (i.e., trapped) situation, we use the ratios of the cloud radii provided by experimental measurements in a trapped imbalanced Fermi gas by Shin et al~\cite{Shin2008} in order to determine $\frac{h_c}{\mu}(r)$ consistently, where $r$ is the position in the trap. We set $\hbar=1$ throughout this paper.

\section{Equilibrium in Homogeneous Systems}

The position of the phase transition from the superfluid to the normal phase can be found analytically, by imposing that the chemical potentials and pressures in the superfluid and normal phases match~\cite{Bedaque2003,Caldas2004,Sanjay}. The mean-field (MF) pressure of an ideal partially polarized normal state, is given by

\begin{eqnarray}
\label{PN}
P^{\rm N}(\mu_{\downarrow\uparrow})&=&\frac{k_{F\uparrow}^5}{30 \pi^2 m} + \frac{k_{F\downarrow}^5}{30 \pi^2 m}
\\
\nonumber
&=& \frac{1}{15 \pi^2} (2m)^{3/2} \left[\mu_{\uparrow}^{5/2} + \mu_{\downarrow}^{5/2} \right]
\\
\nonumber
&=& \frac{(2m)^{3/2}}{15 \pi^2}  \mu^{5/2} \left[\left(1+ \frac{h}{\mu} \right)^{5/2} + \left(1- \frac{h}{\mu} \right)^{5/2} \right],
\end{eqnarray}
where $\mu_{\uparrow} = \mu + h$, and $\mu_{\downarrow} = \mu - h$ are the chemical potentials of the spin-up and spin-down particles, and $m$ is the fermion mass.

Universality at the unitary limit allows one to write the pressure in the superfluid phase (SP) as~\cite{ChevyPRL,ChevyPRA}

\begin{eqnarray}
\label{PS}
P^{\rm S}(\mu)&=& \frac{1}{15 \pi^2} \left(\frac{m}{\xi}\right)^{3/2}  \left(\mu_{\uparrow} + \mu_{\downarrow} \right)^{5/2} 
\\
\nonumber
&=& \frac{1}{15 \pi^2} \left(\frac{m}{\xi}\right)^{3/2} \left(2\mu \right)^{5/2},
\end{eqnarray}
where $\xi$ is an universal parameter, which is obtained from theoretical (quantum Monte Carlo (QMC)) calculations~\cite{Carlson,Perali,Boronat,Sanjay}, as well as from experiments~\cite{Exp2,Ohara,Bartenstein,Kinast}.

The equilibrium between the two phases is reached when $P^{\rm N}(\mu_{\downarrow\uparrow})=P^{\rm S}(\mu)$, and $\mu=(\mu_{\uparrow} + \mu_{\downarrow})/2$~\cite{Bedaque2003,Caldas2004}, which gives

\begin{eqnarray}
\label{PN-PS}
\left(1+ \frac{h}{\mu} \right)^{5/2} + \left(1- \frac{h}{\mu} \right)^{5/2} = \frac{2}{\xi^{3/2}}.
\end{eqnarray}
As pointed out in Ref.~\cite{He}, the authors of Ref.~\cite{Sanjay} assumed that the transition is between the SP and the fully FP normal phase, which implies in setting $h=\mu$ in Eq.~(\ref{PN-PS}), yielding

\begin{eqnarray}
\label{PN-PS2}
\frac{h_c}{\mu} = \frac{2^{2/5}}{\xi^{3/5}}  -1.
\end{eqnarray}
On the other hand, at the universal regime the pairing gap is $\Delta = \beta \mu$, where QMC results give $\xi \sim 0.42(1)$ and $\beta \sim 1.2(1)$.  Then, they found~\cite{Sanjay}

\begin{eqnarray}
\label{PN-PS3}
\frac{h_c}{\Delta} = \frac{1}{\beta}  \left(\frac{2^{2/5}}{\xi^{3/5}}  -1\right) = 1.00(5).
\end{eqnarray}

However, one striking feature of Ref.~\cite{Exp1} is the observation of three different regions (phases) in the cloud. At the center, they observed a superfluid core, where the densities of the two spin states are equal, then an intermediate normal shell, containing equal densities of the two spin states, and finally an outer region, with only atoms of the majority component.

Based on this fact, we now consider both populations in the {\it intermediate} normal phase, and expand Eq.~(\ref{PN-PS}) up to second order in $h/\mu$ and combine this result with the gap at the universal regime to find the ratio

\begin{eqnarray}
\frac{h_c}{\Delta} = \frac{1}{\beta} \sqrt{\frac{8}{15}} \left(\frac{1}{\xi^{3/2}} -1 \right)^{1/2}.
\end{eqnarray}
Plugging the values of $\xi$ and $\beta$ into the above equation we find $h_c/\Delta \simeq 0.81$.

However, this result is not reliable since this equation gives an wrong result for the maximum chemical potential imbalance

\begin{eqnarray}
\frac{h_c}{\mu} = \sqrt{\frac{8}{15}} \left(\frac{1}{\xi^{3/2}} -1 \right)^{1/2} \sim 1.138.
\end{eqnarray}
This result is wrong by three reasons: \\
{\bf i)} it was obtained under the assumption that $h/\mu \ll 1$, and the value found is $>1$;\\
{\bf ii)} this $h_c > h_{max}$, where $h_{max}$ is the maximum value for the MF pressure in Eq.~(\ref{PN}), which is $h_{max}=\mu$ such that $h_{max}/\mu=1$ (meaning that from Eq.~(\ref{PN}) the normal phase is fully polarized at $h_{max}$);\\
{\bf iii)} it was obtained from expression~(\ref{PN}) which completely neglects interactions in the normal phase.

Then, we have seen that the problem in obtaining the correct critical chemical polarization at which the $SP-N$ phase transition occurs is not only due to do not taking into account both species in the normal phase as in Eq.~(\ref{PN-PS2}), but also, and mainly, due to the fact that the superfluid pressure $P^{\rm S}(\mu)$ in Eq.~(\ref{PS}) is valid at unitarity, while $P^{\rm N}(\mu_{\downarrow\uparrow})$ in Eq.~(\ref{PN}) is the pressure of an ideal (not-interacting) imbalanced Fermi gas.

Let us now take the MF pressure in the BCS state 

\begin{equation}
P^{\rm BCS}( \mu)=\frac{k_{\rm
  F}^5}{ 15 \pi^2 m} + \frac{mk_{\rm F}}{4\pi^2} \Delta_0^2,
\label{BCS}
\end{equation}
where $\Delta_0=\left( \frac{2}{e}\right)^{7/3}\mu e^{-\pi/2k_F|a|}$ is the BCS pairing gap. Equating the pressures of the BCS and normal state we find

\begin{eqnarray}
\label{CC-0}
&&\left(1+\frac{h}{\mu} \right)^{\frac{5}{2}} + \left(1- \frac{h}{\mu}\right)^{\frac{5}{2}} = 2 + \frac{30}{16}\frac{\Delta_0^2}{\mu^2}.
\end{eqnarray}
As before, we expand the above equation up to second order in ${h}/{\mu}$, which gives

\begin{eqnarray}
\label{CC-1}
\frac{15}{4} \left(\frac{h}{\mu} \right)^2 = \frac{30}{16}\frac{\Delta_0^2}{\mu^2},
\end{eqnarray}
whose solution gives the well known Chandrasekhar-Clogston (CC) limit of superfluidity

\begin{eqnarray}
\label{CC}
\frac{h_{cc}}{\Delta_0}=\frac{1}{\sqrt{2}}.
\end{eqnarray}
Therefore, the weak-coupling MF solution is consistent and valid.

\section{Effect of Interactions in the Normal Phase}

One of the main effects of interactions is the modification of the chemical potentials of the spin-$\uparrow$ and spin-$\downarrow$ species. The failure of Eq.~(\ref{PN}) in describing the SF-N phase transition is due to the fact that this equation can not be used beyond the weak-coupling MF limit, as in Eq.~(\ref{CC}). To have a correct description of the phase transition in the strong coupling limit and mainly, in the unitary limit, the chemical potentials have to be modified accordingly in order to take into account the effects of the interactions. In other words, {\it the chemical potentials $\mu_{\uparrow}$ and $\mu_{\downarrow}$ in Eq.~(\ref{PN}) are the initial ones, which do not depend on the interactions}.

We have just seen that at the unitary limit $\mu=\xi E_F$, where $E_F$ is the Fermi energy of a noninteracting gas of density $n=n_{\uparrow}=n_{\downarrow}$, and $\xi$ is the universal parameter we have introduced before.

We follow Ref.~\cite{Exp3,Mora} and write the equation of the pressure of a normal mixed (N-M) state, as a sum of the pressures of an ideal gas of majority atoms with chemical potential $\mu_{\uparrow}$ and an ideal gas of polarons with chemical potential given at unitarity by $\mu_p = A\mu_{\uparrow}$, with $A=-0.61$,

\begin{widetext}
\begin{eqnarray}
\label{PNM}
P^{\rm N-M}(\mu_{\downarrow\uparrow}) &=& \frac{1}{15 \pi^2} (2m)^{3/2} \mu_{\uparrow}^{5/2} + \frac{1}{15 \pi^2} (2m^*)^{3/2} (\mu_{\downarrow}-\mu_{p})^{5/2},
\\
\nonumber
&=& \frac{1}{15 \pi^2} (2m)^{3/2} \mu_{\uparrow}^{5/2} \left[1 + \left(\frac{m^*}{m} \right)^{3/2} (\eta + |A|)^{5/2} \right],
\end{eqnarray}
\end{widetext}
where $m^*$ is the effective polaron mass, and $\eta \equiv \mu_{\downarrow}/ \mu_{\uparrow}$.

We rewrite Eq.~(\ref{PS}) as

\begin{eqnarray}
\label{PS2}
P^{\rm S}(\mu)&=& \frac{1}{15 \pi^2} \left(\frac{m}{\xi}\right)^{3/2}  \left(\mu_{\uparrow} + \mu_{\downarrow} \right)^{5/2} 
\\
\nonumber
&=& \frac{1}{15 \pi^2} \left(\frac{2m}{2\xi}\right)^{3/2} \mu_{\uparrow}^{5/2} \left(1+\eta \right)^{5/2}.
\end{eqnarray}
The equilibrium conditions ($\mu=(\mu_{\uparrow} + \mu_{\downarrow})/2$ and $P^{\rm S}(\mu)= P^{\rm N-M}(\mu_{\downarrow\uparrow}))$ between the N-M and S phase give

\begin{eqnarray}
\label{PS3}
1 + \left(\frac{m^*}{m} \right)^{3/2} (\eta + |A|)^{5/2}=\frac{1}{\left(2\xi \right)^{3/2}} \left(1+\eta \right)^{5/2}.
\end{eqnarray}
We solve this equation for $m^*=1.22m$ and find

\begin{eqnarray}
\label{Equilibrium3}
\eta_{c1} \sim 0.061,
\end{eqnarray}
which is near the experimental value $\eta_{c1} \sim 0.065$, Ref.~\cite{Exp3}. 

The ratio $\eta$ of the chemical potentials can be written as

\begin{eqnarray}
\label{ratioeta1}
\eta = \frac{\mu_{\downarrow}}{\mu_{\uparrow}}=\frac{1-h/\mu}{1+h/\mu}.
\end{eqnarray}
Notice that the $\eta_{c}$ correspondent to the $h_{c}/\mu=\frac{2^{2/5}}{\xi^{3/5}}  -1$ in Eq.~(\ref{PN-PS2}) is given by using the above equation by $\eta_c = (2\xi)^{3/5} -1 \sim - 0.099$. However, as we have mentioned before, this value of $h_{c}/\mu$ (and consequently $\eta_c$) corresponds to a transition from the superfluid phase to a fully polarized normal phase, which is not the situation found in experiments~\cite{Exp1}.

From Eq.~(\ref{ratioeta1}) one quickly finds

\begin{eqnarray}
\label{saturation1}
\frac{h}{\mu} = \frac{1-\eta}{1+\eta}.
\end{eqnarray}

This equation allows us to obtain $\delta \mu_{c1}/\mu=h_{c1}/\mu$ as a function of $\eta_{c1}$ from Eq.~(\ref{Equilibrium3}),

\begin{eqnarray}
\label{PS4}
\left(\frac{\delta \mu}{\mu} \right)_c = 0.88,
\end{eqnarray}
which is between the values found by Frank et al.~\cite{Frank}, which obtained $(\delta \mu/\mu)_c=1.09$, Lobo et al.~\cite{Lobo}, that found $(\delta \mu/\mu)_c=0.96$, and Boettcher et al., which found  $(\delta \mu/\mu)_c=0.83$~\cite{Boettcher}. The $(\delta \mu/\mu)_c$ from Eq.~(\ref{PS4}) yields,

\begin{eqnarray}
\label{PS5}
\frac{\delta \mu_c}{\Delta} = \frac{0.88}{\beta} \approx 0.74,
\end{eqnarray}
where the MF result from Eq.~(\ref{CC}) gives $\frac{\delta \mu_c}{\Delta} \approx 0.71$. There is also the investigation of the ratios $\left(\frac{\delta \mu}{\mu} \right)_c$ and $\frac{\delta \mu_c}{\Delta}$ with MF plus first order of $k_Fa$ corrections~\cite{CaldasArxiv}.

The $\delta \mu_c$ found above sets the transition from the superfluid to a normal mixed or, as we have mentioned earlier, intermediate phase. There is another phase transition that occurs from this normal polarized (mixed) to a fully polarized (FP) normal phase, that we now discuss. The pressure of the (noninteracting) FP normal phase is given by

\begin{eqnarray}
\label{FP}
P^{\rm FP}(\mu_{\uparrow}) &=& \frac{1}{15 \pi^2} (2m)^{3/2} \mu_{\uparrow}^{5/2}.
\end{eqnarray}
Equating the pressures of the N-M and N-FP normal phases, we immediately find

\begin{eqnarray}
\label{equi2}
1 + \left(\frac{m^*}{m} \right)^{3/2} (\eta + |A|)^{5/2} =1,
\end{eqnarray}
which has a (unique) solution

\begin{eqnarray}
\label{equi3}
\eta_{c2} = - |A| = -0.615,
\end{eqnarray}
which is in agreement with the interval obtained by Chevy in Ref.~\cite{ChevyPRA} by a different manner: $-0.62 <  \eta_{c2} < -0.61$. The ratio $\eta_{c2}$ fixes the corresponding saturation field $(h/\mu)_{s}$~\cite{Frank}

\begin{eqnarray}
\label{saturation2}
\left(\frac{h}{\mu} \right)_{s} = \frac{1-\eta_{c2}}{1+\eta_{c2}} = 4.19.
\end{eqnarray}

\section{Equilibrium in Trapped Systems}

As we have seen in the previous section, the superfluid-normal and mixed-normal-fully polarized quantum phase transitions are driven by the ratio $h/\mu$. This ratio has been inferred from in situ imaging of the density profiles in a trapped imbalanced Fermi gas by Shin et al~\cite{Shin2008}.

Thus, since experiments take place in the presence of a trapping potential $V(r)$, we want to investigate the effects of the trap in these phase transitions. To account for the effect of an external harmonic confining potential, the system is taken to be locally uniform within the local density approximation (LDA), with a spatially varying local chemical potential, which is given by

\begin{eqnarray}
\label{chemical1}
\mu_{\uparrow}(r) &=& \mu_{\uparrow} - V(r),\\
\nonumber
\mu_{\downarrow}(r) &=& \mu_{\downarrow} - V(r),
\end{eqnarray}
where $\mu_{\uparrow \downarrow} = (\mu \pm h)/2$ are the global chemical potentials we defined before, and we consider a spherical trap without loss of generality $V(r)=\frac{1}{2} m \omega^2 r^2$. Here $\omega$ is the radial trap frequency, and $r$ is the spatial extent of the cloud in the radial direction. 

It is interesting to notice that the average chemical potential is independent of the ``field'' $h$ but varies (with respect to the position $r$) across the trap

\begin{eqnarray}
\label{chemical2}
\mu(r)=\frac{\mu_{\uparrow}(r) + \mu_{\downarrow}(r)}{2}= \mu - V(r),
\end{eqnarray}
where $\mu = (\mu_{\uparrow} + \mu_{\downarrow})/2$, while the chemical potential difference

\begin{eqnarray}
\label{chemical3}
\delta \mu(r) = \delta \mu(r=0) = 2h,
\end{eqnarray}
is constant throughout the trap, but depends on $h$.

The concentric spheres with the different phases have the following radii {\it hierarchy} $R_{SF} < R_{\downarrow} <  R_{\uparrow}$~\cite{PhaseSep,Shin2008}. For $r<R_{SF}$ the gas is in the superfluid (fully paired) state, $R_{SF}<r<R_{\downarrow}$ is the rim of the normal intermediate phase, and $R_{\downarrow} < r < R_{\uparrow}$ is the region where the gas is fully polarized.

The critical radius $R_{SF}$ is determined through the CC criterion $\Delta(R_{SF})= \left( \frac{2}{e}\right)^{7/3}\mu(R_{SF}) e^{-\pi/2k_F(R_{SF})|a|} = \sqrt{2}h$. We approximate the solution for $R_{SF}$ neglecting the dependence of the local Fermi wave vector $k_F(R_{SF})=\sqrt{2m \mu(R_{SF})}$ on $R_{SF}$, and find

\begin{eqnarray}
R_{SF}=\sqrt{1-\frac{h}{h_{cc}}}R_{TF},
\label{chemical4}
\end{eqnarray}
where $h_{cc}$ is given by Eq.~(\ref{CC}), and $R_{TF}=\sqrt{\frac{2\mu}{m \omega^2}}$ is the Thomas-Fermi radius. The condition $\mu_{\downarrow\uparrow}(r=R_{\downarrow\uparrow}) = \mu_{\downarrow\uparrow} - V(r=R_{\downarrow\uparrow}) =0$ defines the Thomas-Fermi radii $R_{TF \downarrow\uparrow}$ of the two species,

\begin{eqnarray}
R_{TF \downarrow}= \sqrt{\frac{2\mu_{\downarrow}}{m\omega^2}}=\sqrt{1-\frac{h}{\mu}} R_{TF},
\label{chemical5}
\end{eqnarray}

\begin{eqnarray}
R_{TF \uparrow}= \sqrt{\frac{2\mu_{\uparrow}}{m\omega^2}}=\sqrt{1+\frac{h}{\mu}} R_{TF}.
\label{chemical6}
\end{eqnarray}
Besides, a simple calculation shows that $R_{TF \uparrow}= \sqrt{R_{TF \downarrow}^2 + \frac{2h}{\mu}R_{TF}^2}$.

The chemical potential ratio in the trap is given by~\cite{Shin2008}

\begin{eqnarray}
\eta(r) &=& \frac{\mu_\downarrow(r)}{\mu_\uparrow(r)} = \frac{\eta \mu_\uparrow - V(r)}{\mu_\uparrow-V(r)} = \frac{\eta - r^2/R_\uparrow^2}{1 - r^2/R_\uparrow^2}\\
\nonumber
&=& 1+ \frac{\eta - 1}{1 - r^2/R_\uparrow^2}.
\label{chemicalratio1}
\end{eqnarray}
We have that $\eta = \frac{ \mu_\downarrow}{\mu_\uparrow} = 2 \frac{ \mu}{\mu_\uparrow} - 1$, where we have used that the thermodynamic equilibrium condition requires $\mu = (\mu_\downarrow + \mu_\uparrow)/2$. Then we have

\begin{eqnarray}
\eta(r) &=& 1+ 2\frac{\mu/\mu_\uparrow - 1}{1 - r^2/R_\uparrow^2}.
\label{chemicalratio2}
\end{eqnarray}
The global chemical potential $\mu$ of a fully-paired superfluid in the core is given $\mu = \xi \epsilon_F= \xi (6\pi^2 n_s)^{2/3}/2m$, where $\epsilon_F$ is the local Fermi energy and $n_s$ is the majority (or minority) density at the center of the trap. Besides, $\mu_\uparrow = (6\pi^2 n_0)^{2/3}/2m$. This finally yields

\begin{eqnarray}
\eta(x) &=& 1+ 2\frac{\xi(n_s/n_0)^{2/3} - 1}{1 - x^2},
\label{chemicalratio3}
\end{eqnarray}
where $x \equiv r/R_\uparrow$.

\begin{figure}[htb]
  \vspace{0.1cm}
 \epsfig{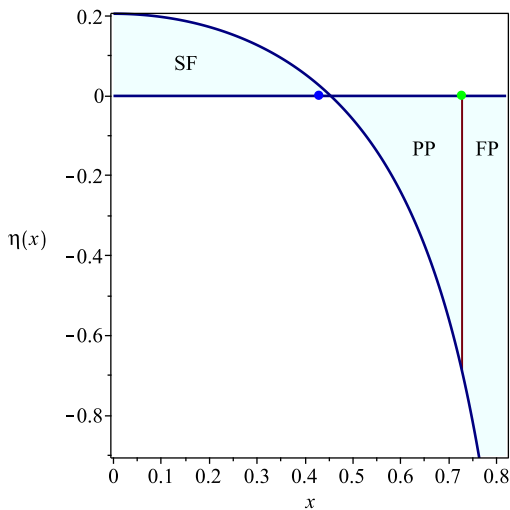}
\caption[]{\label{eta} (Color online) The chemical potential ratio $\eta(x)$ as a function of the ``normalized'' position in the trap $x=r/R_\uparrow$. The blue point corresponds to the position $R_c/R_\uparrow=0.43$, while the green one refers to $R_\downarrow/R_\uparrow = 0.728$.}
\end{figure}

The measured values~\cite{Shin2008} for the ratio of the radius $R_c$ of the balanced superfluid in the center of the trap to the outer radius $R_\uparrow$ of the fully polarized gas at the edge is $R_c/R_\uparrow=0.43$, and the normalized central density $n_s/n_0 = 1.72$, with $\xi = 0.42$ Eq.~(\ref{chemicalratio3}) gives $\eta_c = \eta(R_c/R_\uparrow) = 0.03$. 

In Fig.~\ref{eta} we show the behavior of $\eta(x)$ as a function of $x$ for $\xi = 0.42$ and $n_s/n_0 = 1.72$. The region between zero and the blue point is the region in the trap with the balanced superfluid (SF). The region between the blue and green point corresponds to the partially polarized (PP) normal phase, while the green and some another point (not shown) refers to the fully polarized (FP) region.

The condition $\eta(x)=0$ yields the value of $x$ at which the ratio $\frac{\mu_\downarrow(r)}{\mu_\uparrow(r)}$ changes sign along the (radial) position in the trap. Notice that with these measured values of $\xi$ and $n_s/n_0$, $\eta(x)=0$ at $x=\sqrt{2\xi(n_s/n_0)^{2/3} - 1} \approx 0.45$ which is close to the measured value $R_c/R_\uparrow=0.43$. Besides, in order this point to exist it is necessary to have $\xi(n_s/n_0)^{2/3} \geq 1/2$ i.e., $\mu/\mu_{\uparrow} \geq 1/2$.

With Eq.~(\ref{saturation1}) it was found~\cite{Shin2008}

\begin{eqnarray}
\left(\frac{h}{\mu} \right)_c = 0.95.
\label{chemicalratio4}
\end{eqnarray}
The measured ratio of the minority to the majority radii $R_\downarrow/R_\uparrow = 0.728$, yielding $\eta_\downarrow = \eta(R_\downarrow/R_\uparrow) \approx - 0.69$, which gives a saturation field

\begin{eqnarray}
\left(\frac{h}{\mu} \right)_s = 5.5.
\label{chemicalratio5}
\end{eqnarray}
However, Eq.~(\ref{saturation1}) was derived for the homogeneous situation, $V(r)=0$. In order to derive an equation for $h/\mu$ which depends on the position $r$ in the trap, we go back to the definition of $\eta(r)$, and derive an equation for $h/\mu$ which is $r$-dependent

\begin{eqnarray}
\eta(r) &=& \frac{\mu_\downarrow(r)}{\mu_\uparrow(r)} = \frac{\mu -h - V(r)}{\mu_\uparrow-V(r)} \\
\nonumber
&=& \frac{\mu(1- h/\mu - r^2/R_{TF}^2)}{\mu_{\uparrow}(1 - r^2/R_\uparrow^2)}.
\label{chemicalr1}
\end{eqnarray}
Inverting the equation above we find $\frac{h}{\mu}=\frac{h}{\mu}(x,\eta)$, which is given by

\begin{eqnarray}
\label{chemicalr2}
\frac{h}{\mu}(x,\eta) = 1 -  \frac{R_\uparrow^2}{R_{TF}^2} x^2 
- \frac{\eta(x)}{\xi(n_s/n_0)^{2/3}} \left( 1 -  x^2\right),
\end{eqnarray}
with $\eta(x)$ given by Eq.~(\ref{chemicalratio3}).

With the values used before for $R_c/R_\uparrow=0.43$, $\eta(R_c/R_\uparrow)=0.03$, $\xi=0.42$, $n_s/n_0=1.72$, and also the measured value of $R_\uparrow/R_{TF} \approx 0.95$~\cite{Shin2008}, from Eq.~(\ref{chemicalr2}) we find

\begin{eqnarray}
\left(\frac{h}{\mu} \right)_c \simeq 0.79,
\label{chemicalr3}
\end{eqnarray}
while for $R_\downarrow/R_\uparrow = 0.728$, and $\eta_\downarrow = \eta(R_\downarrow/R_\uparrow) \approx - 0.69$, we obtain

\begin{eqnarray}
\left(\frac{h}{\mu} \right)_s \simeq 1.059.
\label{chemicalr3}
\end{eqnarray}
Notice that if one would like to know what would be the result of Eq.~(\ref{chemicalr2}) if that equation did not depend on $r$, we take $x=0$ in Eq.~(\ref{chemicalr2}), which gives $\left(\frac{h}{\mu}(x=0,\eta)\right) _c = 1- \frac{\eta}{\xi(n_s/n_0)^{2/3}}$ that, for $\eta=0.03$, yields $\left(\frac{h}{\mu}(x=0,\eta=0.03)\right) _c \approx 0.95$, which is the result in Eq.~(\ref{chemicalratio4}). However, for $\eta=-0.69$ we find $\left(\frac{h}{\mu}(x=0,\eta= -0.69)\right) _s \approx 2.14$, which is a much lower value than the one in Eq.~(\ref{chemicalratio5}).

To find the transition from FP phase, of the majority species with boundary at $R_\uparrow$, and the vacuum, with no particles, we insert Eq.~(\ref{chemicalratio3}) into Eq.~(\ref{chemicalr2}), and obtain

\begin{eqnarray}
\label{hmax}
\left(\frac{h}{\mu} \right)_{m}&\equiv& \frac{h}{\mu}(x=1) = -1 -  \frac{R_\uparrow^2}{R_{TF}^2} + \frac{2}{\xi(n_s/n_0)^{2/3}} \\
\nonumber
&\simeq& 1.41.
\end{eqnarray}
In homogeneous systems, the transition to vacuum is given by $h_{m}=|\mu|$~\cite{Sandro,Frank}.

\begin{figure}[htb]
  \vspace{0.1cm}
  \epsfig{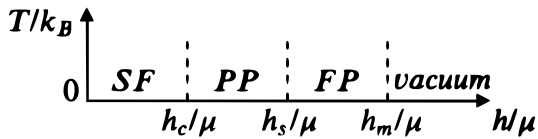}
\caption[]{\label{ZT} The zero temperature phase diagram of a trapped imbalanced Fermi gas at unitarity, as seen in experiments~\cite{Shin2008}.}
\end{figure}

In Fig.~\ref{ZT} we show the sketch of the zero temperature phase diagram of the trapped imbalanced Fermi gas at unitarity. The ``effective'' fields and order of the quantum phase transitions are $h_c/\mu = 0.79$, a first-order phase transition from the unpolarized SF to a PP normal phase, $h_s/\mu = 1.06$, a second-order phase transition from the PP to the FP normal phase~\cite{Shin2008,Sandro}, and the maximum field which sets the transition from FP normal phase to the vacuum, $h_m/\mu= 1.41$, which is also of second-order.

\begin{table}[h!]
  \begin{center}
    \caption{Critical chemical potential imbalance $(h/\mu)_c$, for the superfluid-normal quantum phase transition at unitarity, calculated with $\xi=0.42$. The acronym SPM stands for simple polaron model, employed in the homogeneous case, and DFE is data from experiments, for the trapped situation. FRG is functional renormalization group, and LW is Luttinger-Ward.}
    
    \label{tab:table1}
    \begin{tabular}{l|c|r} 
      \textbf{Work} & \textbf{Homegeneous} & \textbf{Trapped}\\
      \hline
     1/N expansion \cite{Sachdev} & 0.808 & \\
    FRG  \cite{Boettcher} & 0.83 & \\
     This work, SPM and DFE & 0.88 & 0.79\\
     QMC \cite{Lobo} & 0.96 & \\
      $\epsilon=4-d$ expansion \cite{Son} & 1.15 &\\
      QMC \cite{Sanjay} & 1.22 & \\
    LW formalism  \cite{Frank} &  & 1.09 \\
     Experiments \cite{Shin2008} & & 0.95 \\
     Experiments \cite{Exp3} & & 0.878\\
    \end{tabular}
  \end{center}
\end{table}

\section{Conclusions and Outlook}

We have reinvestigated the various quantum phase transitions that may happen in a cold Fermi gas with imbalanced spin populations, triggered by the ratio of the effective Zeeman magnetic field $h$ to the chemical potential $\mu$, $h/\mu$. The analysis has been performed for both homogeneous and trapped systems, focusing on the phase transitions from superfluid to normal-mixed (or PP), from normal-mixed to FP, and finally, from FP to the vacuum. 

For a comparison, some results from several approaches for the superfluid-normal phase transition are presented in table~\ref{tab:table1}.

It is important to mention that, as pointed out in Ref.~\cite{Frank}, the value of $(h/\mu)_c \approx 0.95$ found for $\eta(R_c/R_\uparrow)=0.03$ was for a Bertsch parameter $\xi = 0.42$, and that for the measured value $\xi = 0.37$~\cite{Ku,Zurn}, Eq.~(\ref{chemicalratio3}) gives $\eta(R_c/R_\uparrow,\xi=0.37)=-0.15$, which by Eq.~(\ref{saturation1}) leads to the much larger critical field $(h/\mu)_c \approx 1.35$. Besides, $\eta(R_\downarrow/R_\uparrow) = -0.99$ for $\xi = 0.37$, gives the value $(h/\mu)_s \approx 400.6$. We find that these huge values are due to the use of $(h/\mu)$ valid for homogeneous systems, Eq.~(\ref{saturation1}), together with $\eta(r/R_\uparrow)$ from Eq.~(\ref{chemicalratio3}), which clearly depends on the position in the trap. 

We have then derived a consistent equation for $h/\mu$ as a function of the ratio $r/R_\uparrow$, Eq.~(\ref{chemicalr2}). Thus, with Eq.~(\ref{chemicalr2}) we find for $\xi = 0.37$, $(h/\mu)_c \approx 1.063$, $(h/\mu)_s \approx 1.4$, and with Eq.~(\ref{hmax}), $(h/\mu)_m= 1.86$, respectively.

As mentioned in Ref.~\cite{Frank}, a very important, and still open issue, is the precise nature of the ground state in the regime $h_c < h < h_s$. As plans for a future work, we intent to investigate the possible phases that may arise in this adverse and, at the same time, intriguing regime, and the respective phase transitions associated to them. One phase which is a potential candidate to appear in this regime is the elusive Fulde-Ferrell-Larkin-Ocvhinnikov (FFLO) state~\cite{Fulde,Larkin,Elusive}.

{\it Acknowledgments:} 

I am grateful to F. Chevy, L. He and W. Zwerger for stimulating conversations. I also wish to thank CNPq and FAPEMIG for partial financial support.


\begin{thebibliography}{10}
\bibitem{Lewenstein} M. Lewenstein, A. Sanpera, V. Ahufinger, and B. Damski, Advances in Physics {\bf 56}(2), 243379 (2007).
\bibitem{Review1} M. Inguscio, W. Ketterle, and C. Salomon. Ultracold Fermi Gases. {\it Proceedings of the International School of Physics Enrico Fermi}, Course CLXIV, Varenna, (2006).
\bibitem{Review2} I. Bloch, J. Dalibard, and W. Zwerger, Rev. Mod. Phys. {\bf 80}(3), 885 (2008).
\bibitem{Teo1} S. Giorgini, L. P. Pitaevskii, and S. Stringari, Rev. Mod. Phys. {\bf 80}, 1215 (2008).
\bibitem{Clogston} A. M. Clogston, Phys. Rev. Lett. {\bf 9}, 266 (1962).
\bibitem{Chandrasekhar} B. S. Chandrasekhar, Appl. Phys. Lett. {\bf 1}, 7 (1962).
\bibitem{Bedaque2003} P.~F. Bedaque, H.~Caldas and G.~Rupak, Phys. Rev. Lett. {\bf 91}, 247002 (2003).
\bibitem{Caldas2004} H.~Caldas, Phys. Rev. A {\bf 69}, 063602 (2004).
\bibitem{Souza} H. Caldas, A. L. Mota, R. L. S. Farias, and L. A. Souza, J. Stat. Mech. (2012) P10019.
\bibitem{Thomas}  W. Ong, C. Cheng, I. Arakelyan, and J. E. Thomas, Phys. Rev. Lett. {\bf 114},110403 (2015).
\bibitem{Mitra} D. Mitra, P. T. Brown, P. Schau§, S. S. Kondov, and W. S. Bakr, Phys. Rev. Lett. {\bf 117}, 093601 (2016).
\bibitem{PhaseSep} Y. Shin, M. Zwierlein, C. Schunck, A. Schirotzek, W. Ketterle, Phys. Rev. Lett. {\bf 97}, 030401 (2006).
\bibitem{Teo2} L. Radzihovsky and D. E Sheehy, Rep. Prog. Phys. {\bf 73}, 076501 (2010).
\bibitem{Teo3} F. Chevy and C. Mora, Rep. Prog. Phys, {\bf 73}, 112401 (2010).
\bibitem{Teo4} W. Zwerger, Ed., {\it The BCS-BEC Crossover and the Unitary Fermi Gas}. vol. 836, Lecture Notes in Physics, Springer, (2012).
\bibitem{Exp1} M. W. Zwierlein et al., Science {\bf 311}, 492 (2006); M. W. Zwierlein et al., Nature (London) {\bf 442}, 54 (2006).
\bibitem{Exp2} G. B. Partridge et al., Phys. Rev. Lett. {\bf 97}, 190407 (2006); G. B. Partridge et al., Science {\bf 311}, 503 (2006).
\bibitem{Exp3} S. Nascimb\`ene, N. Navon, K. J. Jiang, F. Chevy, and C. Salomon, Nature (London) {\bf 463}, 1057 (2010).
\bibitem{Mora} C. Mora and F. Chevy, Phys. Rev. Lett. {\bf 104}, 230402 (2010).
\bibitem{Exp4} N. Navon, S. Nascimbene, F. Chevy, and C. Salomon, Science {\bf 328}, 729 (2010).
\bibitem{Exp5} S. Nascimb\`ene, N. Navon, S. Pilati, F. Chevy, S. Giorgini, A. Georges, and C. Salomon, Phys. Rev. Lett. {\bf 106}, 215303 (2011).
\bibitem{Shin2008} Y. Shin, C. H. Schunck, A. Schirotzek, and W. Ketterle, Nature {\bf 451}, 689 (2008).
\bibitem{QI} A. Osterloh, L. Amico, G. Falci, and R. Fazio, Nature {\bf 416}, 608 (2002).
\bibitem{ChevyPRL} F. Chevy, Phys. Lett. A {\bf 96}, 130401 (2006).
\bibitem{ChevyPRA} F. Chevy, Phys. Rev. A {\bf 74}, 063628 (2006).
\bibitem{He} L. He, and P. Zhuang, Phys. Rev. B {\bf 83}, 174504 (2011).
\bibitem{Carlson} J. Carlson, S.-Y. Chang, V. R. Pandharipande, and K. E. Schmidt, Phys. Rev. Lett. {\bf 91}, 050401 (2003).
\bibitem{Perali} A. Perali, P. Pieri, and G. C. Strinati, Phys. Rev. Lett. {\bf 93}, 100404 (2004).
\bibitem{Boronat} G. E. Astrakharchik, J. Boronat, J. Casulleras, and S. Giorgini, Phys. Rev. Lett. {\bf 93}, 200404 (2004).
\bibitem{Sanjay} J. Carlson and S. Reddy, Phys. Rev. Lett. {\bf 95}, 060401 (2005).
\bibitem{Ohara} K. M. O'Hara, S. L. Hemmer, M. E. Gehm, S. R. Granade, and J. E. Thomas, Science {\bf 298}, 2179 (2002).
\bibitem{Bartenstein} M. Bartenstein et al., Phys. Rev. Lett. {\bf 92}, 120401 (2004).
\bibitem{Kinast} J. Kinast et al., Science {\bf 307}, 1296 (2005).
\bibitem{Lobo} C. Lobo, A. Recati, S. Giorgini, and S. Stringari, Phys. Rev. Lett. {\bf 97}, 200403 (2006).
\bibitem{Sachdev} P. Nikolic and S. Sachdev, Phys. Rev. A {\bf 75}, 033608 (2007).
\bibitem{Son} Y. Nishida and D. T. Son, Phys. Rev. A {\bf 75}, 063617 (2007).
\bibitem{Boettcher} I. Boettcher, J. Braun, T. K. Herbst, et al., Phys. Rev. A {\bf 91}, 013610 (2015).
\bibitem{Frank} B. Frank, J. Lang, and W. Zwerger, JETP {\bf 127}(5), 812 (2018).
\bibitem{CaldasArxiv} H. Caldas, J. Stat. Mech. (2019) P103102.
\bibitem{Sandro} I. Bausmerth, A. Recati, and S. Stringari, Phys. Rev. A {\bf 79}, 043622 (2009).
\bibitem{Ku} M. J. H. Ku {\it et al.} Science {\bf 335}, 563 (2012).
\bibitem{Zurn} G. Zurn, T. Lompe, A. N. Wenz, et al., Phys. Rev. Lett. {\bf 110}, 135301 (2013).
\bibitem{Fulde} Fulde P. and Ferrell R. A., Phys. Rev., {\bf 135} A550 (1964).
\bibitem{Larkin} Larkin A. I. and Ovchinnikov Y. N., Zh. Exp. Teor. Fiz., {\bf 47} 1136 (1964) (Sov. Phys. JETP, {\bf 20} 762 (1965)).
\bibitem{Elusive} J. J. Kinnunen, J. E. Baarsma, J.-P. Martikainen and P. T\"orm\"a, Rep. Prog. Phys. {\bf 81} 046401 (2018).
\end{thebibliography}
\end{document}